\documentclass{PoS}
\usepackage{lineno}
\usepackage{caption}
\usepackage{subcaption}
\usepackage{etoolbox}
\patchcmd{\thebibliography}{\section*{\refname}}{}{}{}

\title{Search for high-energy neutrinos from dust obscured Blazars}

\ShortTitle{Search for neutrinos from obscured Blazars}

\author{\speaker{G. Maggi}$^{1}$, S. Buitink$^{1}$, P. Correa$^{1}$, K.D. de Vries$^{1}$, G. Gentile$^{1,2}$, O. Scholten$^{1,3}$, N. van Eijndhoven$^{1}$\\
\llap{$^1$}Vrije Universiteit Brussel, Pleinlaan 2, 1050 Brussels, Belgium\\
\llap{$^2$}Sterrenkundig Observatorium, Universiteit Gent, Krijgslaan 281, 9000 Gent, Belgium\\
\llap{$^3$}Univ. of Groningen, KVI-Center for Advanced Radiation Technology, Groningen, The Netherlands\\
E-mail: \email{giuliano.maggi@vub.ac.be}}

\abstract{The recent discovery of high-energy cosmic neutrinos by the IceCube neutrino observatory opens up a new field in physics, the field of neutrino astronomy. Using the IceCube neutrino detector we plan to search for high-energy neutrinos emitted from Active Galactic Nuclei (AGN), since AGN are believed to be one of the most promising sources of the most energetic cosmic rays and hence of high-energy neutrinos. We discuss a specific type of AGN which we plan to investigate in more detail with data obtained by the IceCube observatory. The main properties of the AGN category in which we are interested are given by a high-energy jet which is pointing in our line of sight defining a class of AGN, called Blazars, and in particular the ones that are obscured by surrounding dust. The jet-matter interaction is expected to give an increased high-energy neutrino production. The properties of this specific type of AGN are expected to give very distinct features in the electromagnetic spectrum, which are discussed in detail.}

\FullConference{The 34th International Cosmic Ray Conference,\\
		30 July- 6 August, 2015\\
		The Hague, The Netherlands}

\begin{document}

\section{Introduction}\label{uhecr}

\begin{figure}[htb]
 \begin{minipage}[b]{\linewidth}
  \centering\includegraphics[width=0.5\textwidth]{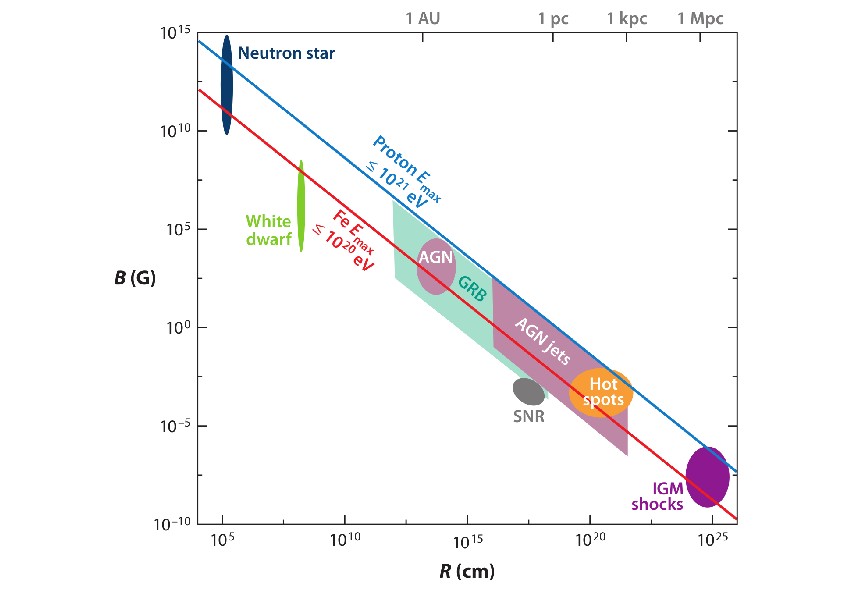}
    \caption{The Hillas diagram, which shows source candidates for UHECRs. Uncertainties are shown on the parameters of the astrophysical candidates. Credit: Kotera \& Olinto \cite{kotera}}
  \label{hillasplot}
 \end{minipage}
\end{figure}

Active Galactic Nuclei (AGN) and Gamma Ray Bursts (GRBs) are the main candidates as sources of ultra-high energy cosmic rays (UHECRs), and even more, UHECRs are believed to originate from extragalactic sources since the galactic magnetic field is not able to contain them. This statement can be exposed by analyzing the Larmor radius $r_{L}= E/ZeB$ $\sim$ $100$ $Z^{-1}$ ($\mathrm{\mu G/B}$)($E/100$ [kpc]) \cite{stanev,kotera}. Furthermore, a relation between magnetic field strength $B$ and extension $R$ of source candidates restricts the maximum energy of the produced cosmic rays as can be seen in the Hillas diagram \cite{hillas}, which is shown in fig. \ref{hillasplot}. This indeed displays that the possible candidates for UHECRs above $10^{18}$ [eV] are GRBs and AGN. The general accepted mechanism for high energy particle production is Fermi shock acceleration, as is outlined in \cite{fermi}.

UHECR detectors, like the Pierre AUGER Observatory \cite{auger} and the Telescope Array Experiment \cite{ta}, have been built in order to detect UHECRs and to search for their origin. With these detectors, UHECRs have already been detected \cite{augerres,tares}, but so far, no correlation between the arrival direction of those particles and astrophysical sources has been found. The main problem for discovering the sources of UHECRs is the existence of magnetic fields in between the source and the detection point, which deflect the trajectory of those charged particles. However, the acceleration of UHECRs implies high-energy neutrino production \cite{julia}. Since neutrinos are chargeless and weakly interacting particles, they can travel through the universe without being affected by magnetic fields and rarely by matter. Those features make high-energy neutrinos the preferred messengers to study the characteristics of UHECR sources.       

The IceCube collaboration already reported High-Energy Extraterrestrial Neutrinos \cite{i3_28}, where the measured flux is close to the Waxman-Bahcall upper bound for the UHE neutrino flux \cite{waxman}. However, so far no specific sources have been identified.

IceCube also reported an upper limit on the flux of high-energy neutrinos associated with GRBs, implying that GRBs can not be the only sources of the most energetic cosmic rays or that the efficiency of neutrino production is much lower than has been predicted \cite{i3_grblimit}. This result strengthens the idea that AGN are good candidates for being the source of high-energy neutrino production.  

In this paper we show our analysis to obtain a class of AGN of special interest. One of the conditions to observe neutrino production from an AGN is that the neutrinos are emitted in the direction of the observer. Therefore, in this work we consider AGN with their jet pointing toward Earth, so-called Blazars. Next to this feature, we want to select a specific type of Blazar. If the jet is interacting with surrounding dust, additional neutrino production is expected from proton-nucleon (p-N) interactions. One of the signatures for dust obscuration of the AGN jet is the attenuation of high-frequency emission in the X-ray and gamma-ray bands. These features will be used to make a selection of dust obscured Blazars from which additional neutrino production is expected in case there is a hadronic component accelerated in the jet. 

\section{AGN classification}\label{agnreview}

An AGN is the luminous compact core of a galaxy, containing a massive Black Hole in its center, with a rotating dust torus on parsec scales. Based on observed characteristics, AGN can be classified in various categories \cite{agnbook}. A definition for different categories of AGN can be found in \cite{agnbook,ned}. Below we will outline the one of interest for our research.
 
\subsection{Blazars}\label{std_Blazars}

A Blazar is observed as a highly variable active galaxy, which in general, displays no emission lines in its spectrum. Futhermore, different studies show that all known Blazars are radio sources \cite{peterson}.
Approximately 10\% of all the AGN are categorized to be a Blazar \cite{mantovani}. Blazars also contain two sub-categories: \emph{flat spectrum radio quasars} (FSRQs) and \emph{BL Lacertae} (BL Lac), depending on the steepness of their radio spectra \cite{agnbook}.

AGN can emit electromagnetic radiation over a wide range of wavelengths, leading to specific signatures in their spectra. In particular, Blazars are well known by being radio-loud AGN, and it is well accepted that the main related physics process taking place in AGN is synchrotron emission \cite{agnbook}. However, radio emission is only one part of the Blazar energy spectrum, which also covers higher frequencies, from X-rays to high-energy $\gamma$-rays. The high-energy spectrum could be explained by two different models, referred as leptonic and hadronic models \cite{leptonic_diltz}, but no clear indication for AGN has been found so far.   

\subsection{AGN obscuration}\label{obs_Blazars}

In the case that there is a cloud of dust in the line of sight between the Blazar and an observer we talk about \emph{AGN obscuration} \cite{andy,bianchi}. The key of this phenomenon is the role that is played by the dust surrounding the Blazar when the particles inside the jet of a Blazar pass through. In this way the dust behaves as an absorbing medium for high-energy emission, for example X-rays, while emission at lower wavelengths is not attenuated. We will call this kind of Blazars, \emph{Dust Obscured Blazars}.     

\section{Neutrino Production in Blazars}

The standard mechanism for producing neutrinos in many high-energy environments is through charged pion decay: $\pi^{+}\rightarrow\nu_{\mu}+\mu^{+}$, $\pi^{-}\rightarrow\bar{\nu_{\mu}}+\mu^{-}$, which would not be the exception in Blazars. If there is a hadronic component in the jet, protons in the jet can interact with the ambient photons producing $p+\gamma\rightarrow\Delta^{+}\rightarrow$$n+\pi^{+}$. This situation would be common for any Blazar.

In the case of Dust Obscured Blazars the protons inside the jet can also interact with matter in the obscuring dust and produce charged pions, which decay in muons and neutrinos. In this way, the neutrino flux would be increased for a Dust Obscured Blazar case, compared to a regular case. 

\begin{figure}[htb]
 \begin{minipage}[b]{\linewidth}
  \centering\includegraphics[width=0.4\textwidth]{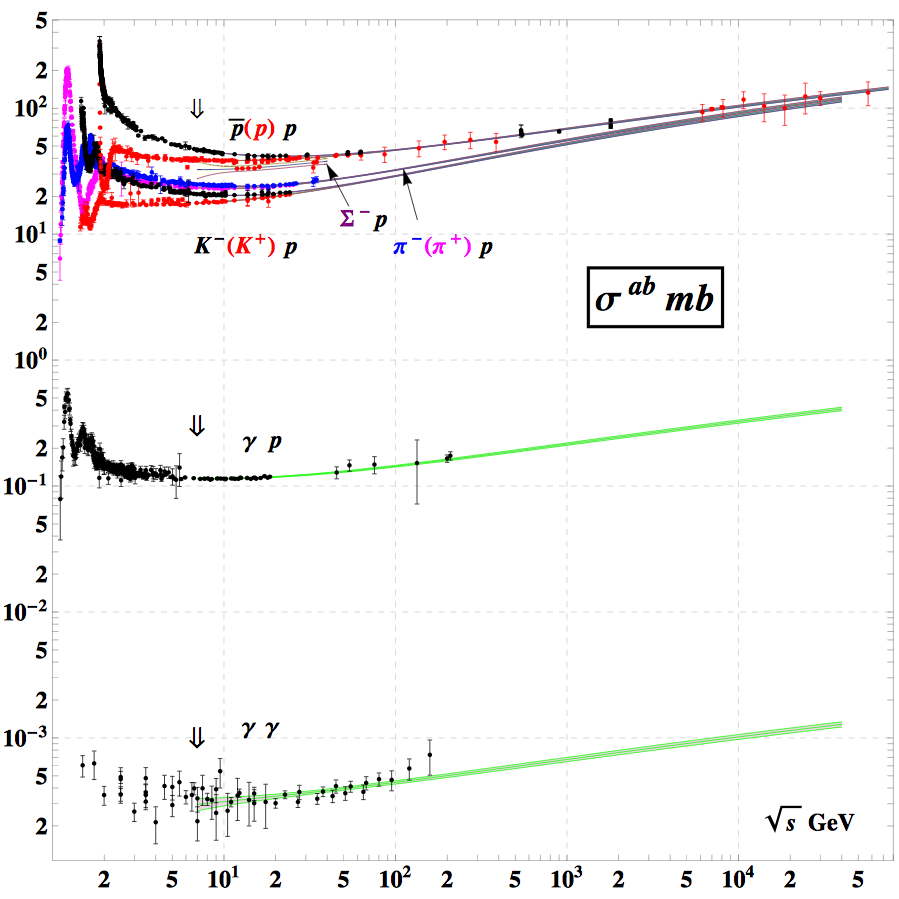}
    \caption{Total cross section $\sigma^{ab}$ in mb for different processes as a function of center of mass energy $\sqrt{s}$ in GeV. The $p+N$ is more than 2 orders of magnitude than $p+\gamma$ cross section \cite{pdg}.}
  \label{xsec}
 \end{minipage}
\end{figure}

We want to emphasize that the cross section for $p+N$ is larger than $p+\gamma$, as is shown in fig. \ref{xsec}. Since the mean free path $\lambda$ is given by $\lambda$$=1/\sigma$$n$, where $n$ is the target density and $\sigma$ is the cross section, a sufficiently dense dust cloud can considerably increase the neutrino yield.   

\section{Neutrino Detection}\label{nu_detection}

A fundamental part of our study is the detection of high-energy neutrinos produced in Blazars. Since neutrinos are only weakly interacting particles, they have a small cross section value compared with other fundamental particle interactions. In fact, some studies show that the largest values for a neutrino cross section would be of the order of $10^{-31}$ cm$^2$ for high-energy neutrinos \cite{quigg}, which still is small compared with other cross sections. Due to their low flux and very small cross section we need a large volume detector to detect astrophysical high-energy neutrinos. For this, the main candidate is IceCube \cite{i3}, which is a particle detector of one cubic kilometer of volume located at the geographic South Pole.

Recentely, the IceCube collaboration reported the discovery of high-energy astrophysical neutrinos \cite{i3_28, i3_37}, which followed up the detection of two PeV neutrino events \cite{i3_2pev}. Our intention is to apply our analysis to IceCube data for up-going muon tracks in the detector, produced by charged current muon-neutrino interaction. Up-going events (northern hemisphere) provide a natural selection in order to avoid atmospheric muon background induced by cosmic rays. 

To analyse the IceCube data we will use a model independent analysis method based on Bayesian statistics to search for signal. In the case that no significant signal is observed, an upper limit is determined. This method is described in \cite{iihepaper} where it was applied to public IceCube data for 10 nearby Blazars, for which a flux upper limit was obtained.  

\begin{table}[htb]
\caption{Selected Point sources from \cite{nijmegen}. The shown values are:  declination (dec), right ascension (ra), $\log$$(\nu$$f_{\nu})$ at $1.4$$\times$$10^{9}$ [Hz] (radio) and $\log$$(\nu$$f_{\nu})$ at $3.02$$\times$$10^{17}$ [Hz] (X-ray) for the 17 selected objects. The empty spots for X-ray measurements are due to the fact that no values where found in the NED catalog \cite{ned} for the flux at $3.02$$\times$$10^{17}$ [Hz]. Possible Dust Obscured Blazar candidates are indicated with a mark (*).}
\begin{center}
        \begin{tabular}{ccccc}\hline
         NED ID                    &   dec       &   ra        &   radio         &   X-ray \\
 \hline
         MRK 0668 *                &   28.45403  &  211.75163  &  9.01901237982  &  9.77835049758 \\
         ARP 220 *                 &   23.50291  &  233.73863  &  8.64443858947  &  9.66809131667 \\
         NGC 0262 *                &   31.95696  &  12.19641   &  8.60097289569  &  10.091730251   \\
         2MASX J16535220+3945369   &   39.76019  &  253.46756  &  9.30998483832  &  12.6779763109 \\
         2MASX J19595975+6508547   &   65.14852  &  299.99924  &  8.50533089667  &  12.5914374947 \\
         IC 0450                   &   74.42698  &  103.05146  &  8.5798978696   &  10.6443597987  \\
         CGCG 204-027              &   41.00279  &  103.79185  &  8.47041049098  &                \\
         NGC 1275                  &   41.51168  &  49.95098   &  10.3662361237  &  13.3550682063  \\
         CGCG 085-029              &   18.58949  &  106.11981  &  8.56276854302  &                \\
         B2 0713+36                &   36.70517  &  109.15459  &  8.64236558084  &                \\
         UGC 03927                 &   59.68421  &  114.3754   &  8.87040390528  &  10.9585734386  \\
         NGC 3801                  &   17.72792  &  175.07042  &  9.19920647916  &                \\
         CGCG 186-048              &   35.01873  &  176.84215  &  8.86003838981  &  10.7027234141 \\
         NGC 3894                  &   59.41562  &  177.20985  &  8.80964050609  &                \\
         NGC 4278                  &   29.2805   &  185.02841  &  8.76511695604  &  10.9189007338  \\
         MCG +10-23-012            &   59.40606  &  239.75703  &  8.45147940512  &                \\
         NGC 6521                  &   62.61216  &  268.95184  &  8.60400993241  &  10.8916266489 \\
         \hline
        \end{tabular}
\end{center}\label{agnselected}
\end{table}

\section{Blazar Selection}

We start our Blazar selection with a recently published catalogue containing 575 radio-emitting galaxies within several hundreds of Mpc \cite{nijmegen}. The radio sources in this catalogue have been separated into four different catagories: starforming galaxies, jets and lobes, unresolved point sources, and unknown morphology. The radio flux measurements were taken from the NVSS \cite{nvss} and SUMMS \cite{sumss} data at 1.4~GHz and 843~MHz respectively. This catalogue was constructed to contain a volume-limited sample of strong radio sources which could be responsible for the UHECRs measured at Earth.

\begin{figure}[htb]
\centering
\begin{subfigure}{.5\textwidth}
  \centering
  \includegraphics[width=.6\linewidth]{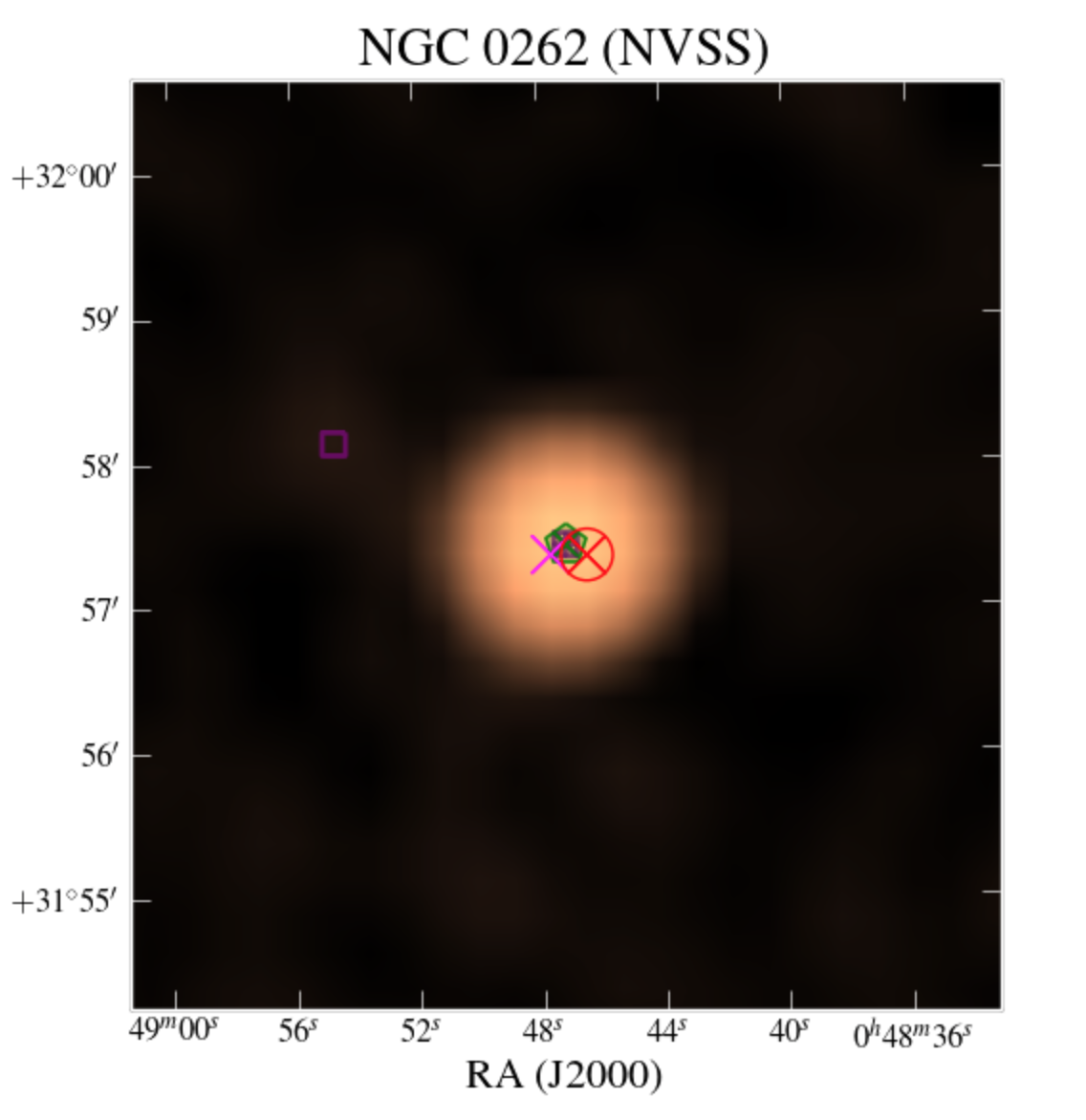}
  \caption{AGN Radio source with a circular morphology.}
  \label{fig:sub1}
\end{subfigure}%
\begin{subfigure}{.5\textwidth}
  \centering
  \includegraphics[width=.6\linewidth]{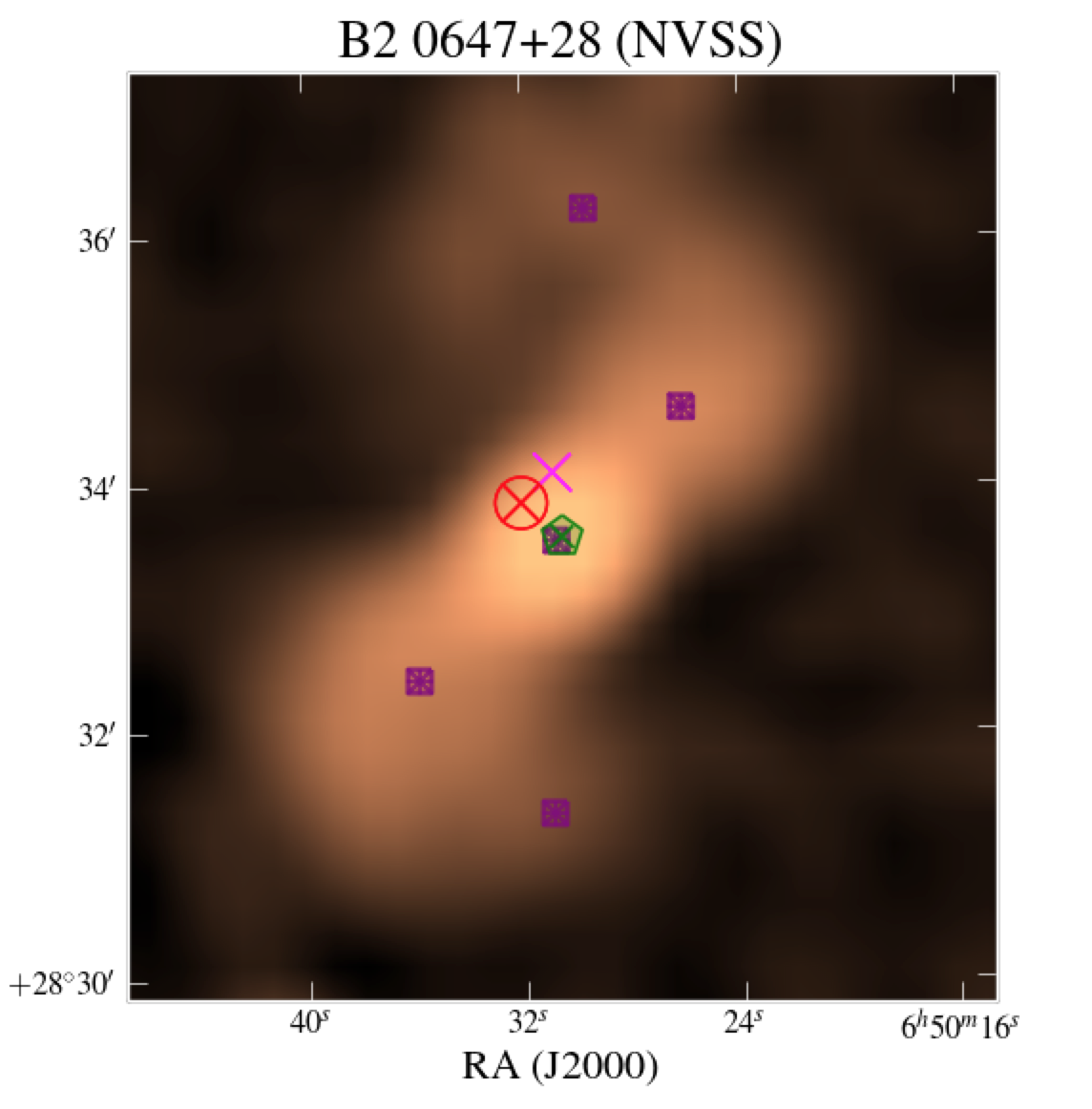}
  \caption{Radio source with a non circular morphology.}
  \label{fig:sub2}
\end{subfigure}
\label{figmorphology}
\end{figure}

Since we plan to use the IceCube neutrino observatory to search for neutrinos emitted from the sources in our final selection, we currently restrict ourselves to sources in the Northern hemisphere. In future work, we plan to apply our selection to sources in the Southern hemisphere as well. After selecting sources in the Northern hemisphere, the next step is to search for Blazars. Since one of the properties of a Blazar is that its jet is pointing toward Earth, a Blazar should have a circular radio morphology. This leaves us with the objects catagorized as point sources. An example is shown in fig. \ref{fig:sub1}, whereas in fig. \ref{fig:sub2} we show one of the objects in the jets and lobes category. After this first selection we are left with 48 interesting objects. 

The next step in our Blazar selection is based on a more quantitative measure. Since the Blazar jet is directed toward Earth, the synchrotron peak in the electromagnetic spectrum of the Blazar will be shifted to higher frequencies. One measure for this shift of the synchrotron peak is the radio spectral index. Therefore, for the remaining objects, the radio spectrum in the range $\nu=$1.4-5~GHz, has been fit by a power law, $F_\nu=C\nu^{-\alpha}$~Jy. The data for this fit is obtained from the Nasa Extragalactic Database \cite{ned}. We selected sources with a radio spectral index $\alpha < 0.5$. These objects are usually classified as Flat Spectrum Radio Quasars (FSRQs), or BL Lac objects. Using this selection, we remain with 17 objects which are listed in table \ref{agnselected}. It should be noted that for 9 point-source objects, due to a lack of data, it was not possible to obtain a radio spectral index and these objects are rejected from our selection.

\begin{figure}[htb]
 \begin{minipage}[b]{\linewidth}
  \centering\includegraphics[width=0.8\textwidth]{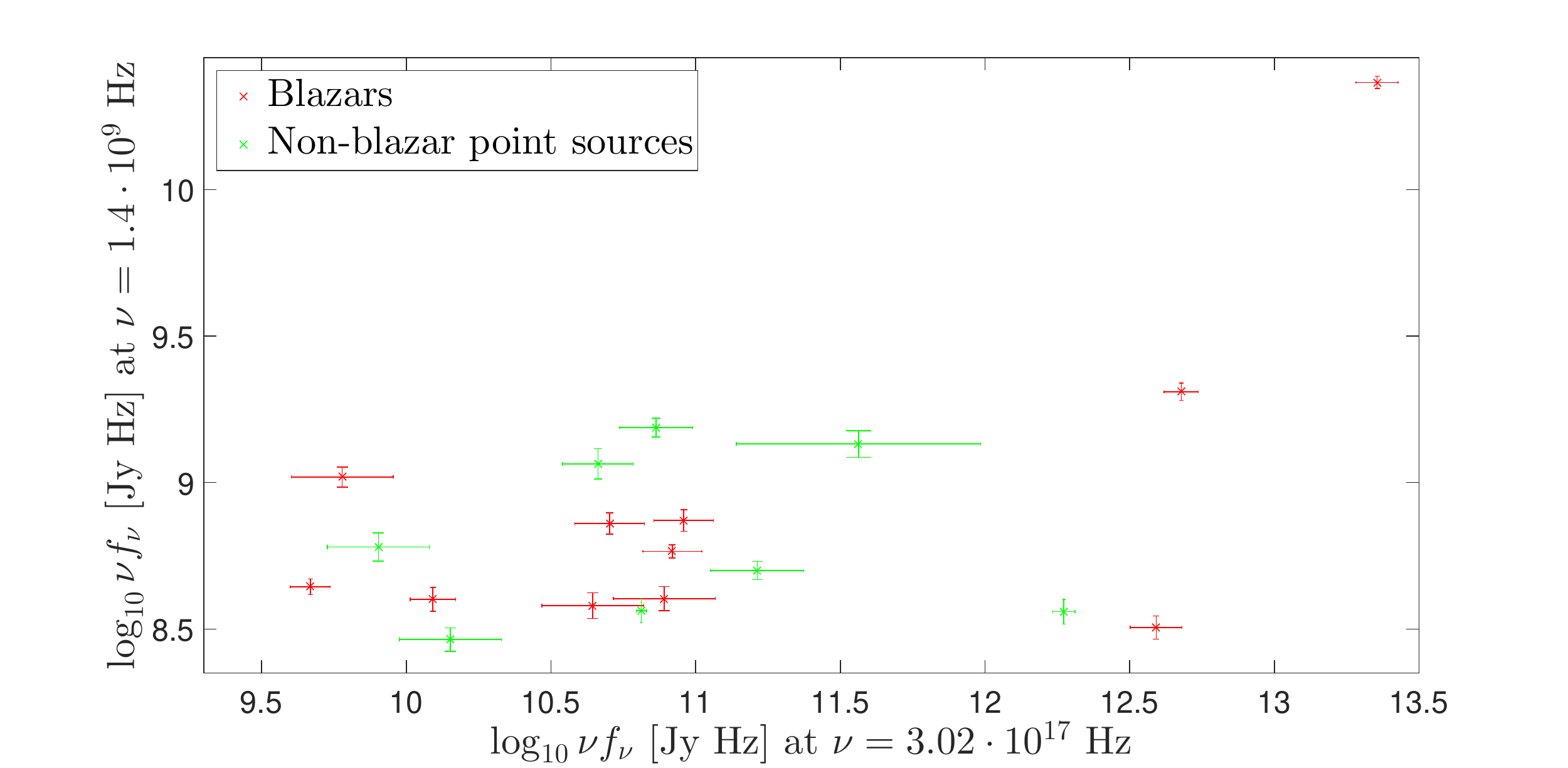}
    \caption{Integrated flux log$(\nu$$f_{\nu})$ at $3.02$$\times$$10^{17}$ Hz for X-rays and $1.4$$\times$$10^{9}$ Hz for radio for our selected objects. Only the objects with radio and X-ray measurement are shown, being 11 out of the 17 initially selected sources.}
  \label{Blazars_north}
 \end{minipage}
\end{figure}

Nevertheless, we are interested in a specific class of Blazars, Dust Obscured Blazars. This leaves us to the final step in our selection. Where emission at small wavelengths in the visible region up to gamma-rays will be attenuated by the dust, emission at longer wavelengths in the radio regime will not be attenuated. Therefore, in fig. \ref{Blazars_north}, we plot the radio flux at 1.4~GHz against the X-ray flux at $3.02\times 10^{17}$~Hz obtained from NED \cite{ned}. It should be noted that for 6 out of the 17 objects no X-ray data could be found in the NED catalogue, and hence these objects might be of great interest for our selection. These objects will be subject of future investigation. In fig. \ref{Blazars_north} we also show the data points for the point-source objects which did not pass our Blazar selection. This is done to see if there is an obvious trend between the radio flux and the X-ray flux. Since no such trend is observed, we finalize our selection by taking a direct cut at an X-ray flux of log$_{10}\nu F_\nu = 10.3$~Jy Hz. This leads us with a final selection of three interesting objects, NGC 0262, ARP 220 and MRK 0668, which could represent Dust Obscured Blazars. 

\section{Summary}
We have presented a selection to search for Dust Obscured Blazars in the Northern hemisphere, where a Blazar is defined by an AGN with its jet directed toward Earth. This specific type of Blazar is expected to give enhanced neutrino production due to the jet-matter interaction in case there is a hadronic component inside the jet and the jet interacts with surrounding dust. Starting from a catalogue of nearby strong radio emitting galaxies, and after several selection procedures based on the radio morphology and detailed analysis of the electromagnetic spectrum, we finally obtain three possible Dust Obscured Blazar candidates, known as: NGC0262, ARP220, and MRK0668. For six objects there was not enough data available to infer whether or not they show features of Dust Obscure Blazar, hence they will be invesigated in more detail in future work.



\begin{thebibliography}{99}
 \bibitem{fermi} E. Fermi. On the Origin of the Cosmic Radiation. Physical Review 75, pp. 1169-1174. (1949).
 \bibitem{hillas} A. Hillas. The Origin of Ultra-High-Energy Cosmic Rays. Annual Review of Astronomy and Astrophysics (1984). 
 \bibitem{stanev} A. Letessier-Selvon and T. Stanev, Ultrahigh Energy Cosmic Rays, Rev. Mod. Phys 83 (2011). 
 \bibitem{kotera} K. Kotera, A. Olinto. The Astrophysics of Ultrahigh Energy Cosmic Rays. Annual Review of Astronomy and Astrophysics vol. 49 (2011).
 \bibitem{boldt} E. Boldt, P. Ghosh. Cosmic rays from remnants of quasars?. Mon.Not.Roy.Astron.Soc.307:491-494, (1999).
 \bibitem{auger} J. Swain (for the Pierre Auger Observatory). AIP Conf.Proc.698:366-369, (2004).
 \bibitem{ta} H. Kaway et al. Telescope Array Experiment. Nuclear Physics B (Proc. Suppl.) 175-176 221-226 (2008).
 \bibitem{augerres} The Pierre Auger Collaboration. Searches for Anisotropies in the Arrival Directions of the Highest Energy Cosmic Rays Detected by the Pierre Auger Observatory. A. Aab et al. 2015 ApJ, 804, (2015).
 \bibitem{tares} Peter Tinyakov for the Telescope Array Collaboration. Latest results from the telescope array. Nuclear Instruments and Methods in Physics Research A 742  29-34 (2014). 
 \bibitem{julia} Julia K. Becker. High-energy neutrinos in the context of multimessenger physics. Phys.Rept.458:173-246,2008. 
 \bibitem{augerref7} V. Berezinsky, G. Zatsepin. Cosmic rays at ultrahigh-energies (neutrino?). Phys.Lett. B28 423-424 (1969).   
 \bibitem{gzk} K. Greisen. End to the Cosmic-Ray Spectrum?. Phys. Rev. Lett. 16, 748, (1966), and V. Zatsepin, V. Kuzmin Upper Limit of the Spectrum of Cosmic Rays. Pisma ZhZhurnal Eksperimental noi i Teoreticheskoi Fiziki 4, 114 (1966). JETP Lett. 4, 78 (1966) (English version).  \bibitem{augerref2} The Pierre Auger Collaboration. Upper limit on the diffuse flux of UHE tau neutrinos from the Pierre Auger Observatory. Phys. Rev. Lett. 100, 211101 (2008). 
 \bibitem{i3_28} IceCube Collaboration. Evidence for High-Energy Extraterrestrial Neutrinos at the IceCube Detector. Science 342, 1242856 (2013).
 \bibitem{i3_37} IceCube Collaboration. Observation of High-Energy Astrophysical Neutrinos in Three Years of IceCube Data. Phys. Rev. Lett. 113, 101101 (2014)
 \bibitem{i3_2pev} IceCube Collaboration. First observation of PeV-energy neutrinos with Ice-Cube. Phys. Rev. Lett. 111, 021103 (2013).
 \bibitem{waxman} E. Waxman, J. Bahcall. High energy astrophysical neutrinos: The upper bound is robust. Phys. Rev. D 64, 023002 (2011).  
 \bibitem{i3_grblimit} IceCube Collaboration. An Absence of Neutrinos Associated with Cosmic Ray Acceleration in Gamma-Ray Bursts. Nature 484, 351-354 (2012).
 \bibitem{agnbook} Hagai Netzer. The Physics and Evolution of Active Galactic Nuclei. Cambridge University Press, (2013).
 \bibitem{ned} NASA/IPAC Extragalactic Database (NED). http://ned.ipac.caltech.edu. 
 \bibitem{peterson} B.M. Peterson. An Introduction to Active Galactic Nuclei. Cambridge University Press, (1997).
 \bibitem{mantovani} F. Mantovani et al. A sample of weak blazars at milli-arcsecond resolution. A\&A 577, A36 (2015). 
 \bibitem{leptonic_diltz} C. Diltz, M. Boettecher.  arXiv:1404.4725, accepted for publication in the Journal of High Energy Astrophysics (JHEAp).
 \bibitem{andy} A. Lawrence, M. Elvis. Misaligned Disc as Obscurers in Active Galaxies. The Astrophysical Journal, Volume 714, Issue 1 (2010).
 \bibitem{bianchi} S. Bianchi, R. Maiolino, G. Risaliti. AGN Obscuration and the Unified Model. Advances in Astronomy, vol. 2012, id. 782030 (2012).
 \bibitem{pdg} J. Beringer et al. Review of Particle Physics. Phys. Rev. D 86, 2012. DOI:10.1103/PhysRevD.86.010001.
 \bibitem{quigg} R. Gandhi, C. Quigg, M. Hall Reno and I. Sarcevic. Ultrahigh-Energy Neutrino Interactions. Astropart.Phys.5:81-110,(1996). 
 \bibitem{i3} A. Karle [IceCube Collaboration]. IceCube - status and recent results. arXiv:1401.4496.
 \bibitem{nijmegen} S. van Velzen et al. Radio galaxies of the local universe: all-sky catalog, luminosity functions, and clustering. A\&A 544, A18 (2012). Also http://ragolu.science.ru.nl .
 \bibitem{iihepaper} D. Bose, L. Brayeur, M. Casier, K. D. de Vries, G. Golup, N. van Eijndhoven. Bayesian Approach for Counting Experiment Statistics applied to a Neutrino Point Source Analysis. Astroparticle Physics 50-52C (2013), pp. 57-64. 
 \bibitem{nvss} The NRAO VLA Sky Survey, Astron. Journ. 115, 1998, p. 1693. 
 \bibitem{sumss} SUMSS. A Wide-Field Radio Imaging Survay of the Southern Sky. II. The Source Catalogue, Mon. Not. Roy Astron. Soc. 342, 2003, p. 1117.
\end{thebibliography}
\end{document}